\documentclass[superscriptaddress,aps,letterpaper,10pt,twocolumn,floats,amsmath,amsfonts,amssymb,pre,nofootinbib]{revtex4-2}
\usepackage{graphicx}
\usepackage[colorlinks=true,citecolor=blue]{hyperref}
\usepackage[caption=false]{subfig}
\usepackage{pstricks}
\usepackage{pst-node}
\usepackage{amsmath}
\usepackage{amsbsy}
\usepackage{mathrsfs}
\usepackage{verbatim}
\usepackage{dsfont}
\usepackage{MnSymbol}
\newcommand{\Tr}{\text{Tr}}

\DeclareMathOperator{\sgn}{sgn}

\begin{document}
	
	
	\title{Macroscopic theory of multipartite correlations in permutation-invariant open quantum systems}
	
	\author{Krzysztof Ptaszy\'{n}ski}
	\email{krzysztof.ptaszynski@ifmpan.poznan.pl}
 	\affiliation{Institute of Molecular Physics, Polish Academy of Sciences, Mariana Smoluchowskiego 17, 60-179 Pozna\'{n}, Poland}

    \affiliation{Complex Systems and Statistical Mechanics, Department of Physics and Materials Science, University of Luxembourg, 30 Avenue des Hauts-Fourneaux, L-4362 Esch-sur-Alzette, Luxembourg}

\author{Maciej Chudak}
 	\affiliation{Institute of Molecular Physics, Polish Academy of Sciences, Mariana Smoluchowskiego 17, 60-179 Pozna\'{n}, Poland}
    
	\author{Massimiliano Esposito}
\affiliation{Complex Systems and Statistical Mechanics, Department of Physics and Materials Science, University of Luxembourg, 30 Avenue des Hauts-Fourneaux, L-4362 Esch-sur-Alzette, Luxembourg}
	
	\date{\today}
	
	\begin{abstract}
Information-theoretic quantities have received significant attention as system-independent measures of correlations in many-body quantum systems, e.g., as universal order parameters of synchronization. In this work, we present a method to determine the macroscopic behavior of the steady-state multipartite mutual information between $N$ interacting units undergoing Markovian evolution that is invariant under unit permutations. Using this approach, we extend a conclusion previously drawn for classical systems that the extensive scaling of mutual information is either not possible for systems relaxing to fixed points of the mean-field dynamics or such scaling is not robust to perturbations of system dynamics. In contrast, robust extensive scaling occurs for system relaxing to time-dependent attractors, e.g., limit cycles.  We illustrate the applicability of our method on the driven-dissipative Lipkin-Meshkov-Glick model.
	\end{abstract}
	
	\maketitle

 \section{Introduction}
 Information theory attracted attention as a universal (system-independent) framework to characterize bipartite and multipartite correlations in many-body quantum systems~\cite{amico2008entanglement,laflorencie2016quantum,de2018genuine}. In particular, a great deal of interest has been focused on the partition size scaling of bipartite correlations between the system partitions in pure~\cite{abanin2019colloquium}, thermal~\cite{wolf2008area}, and nonequilibrium steady states~\cite{gullans2019entanglement,panda2020entanglement,dabruzzo2022logarithmic,caceffo2023entanglement,fraenkel2023extensive,fraenkel2024exact}, with different systems exhibiting either extensive or subextensive scaling. Bipartite correlations have also been used as a signature of quantum~\cite{sarandy2009classical,maziero2010quantum}, thermal~\cite{wilms2012finite} or dissipative~\cite{mattes2023entangled} phase transitions. Finally, mutual information and other information-theoretic measures have been used as universal order parameters of synchronization in coupled oscillator systems~\cite{zhou2002noise,boccaletti2002synchronization,giorgi2012quantum,manzano2013synchronization,giorgi2013spontaneous,ameri2015mutual,zhu2015synchronization,bastidas2015quantum,benedetti2016minimal,davis2016synchronization,galve2017quantum,roulet2018quantum,cardenas2019enhanced,siwiak2020synchronization,jaseem2020generalized,ghosh2022anticipating,ghosh2022early,shen2023fisher, kumar2025understanding}. Recently, increasing research interest has also been devoted to multipartite correlations~\cite{lucke2014detecting,girolami2017quantifying,calegari2020genuine}. For example, their extensive scaling was shown to witness certain quantum~\cite{lourenco2020genuine, lourenço2024genuine, vesperini2024entanglement} and dissipative~\cite{lourenco2022genuine,passarelli2025nonstabilizerness} phase transitions, or ergodicity breaking in disordered quantum systems~\cite{goold2015total,pietracaprina2017total}. They have also been analyzed in the context of work extraction~\cite{braga2014maxwell,yada2024measuring} or witnessing non-Markovian dynamics~\cite{paula2016normarkovianity}.

Characterization of information-theoretic quantities in many-body quantum systems is often cumbersome, as it requires the knowledge of density matrix of the system, whose dimension grows exponentially with system size. The problem can be greatly simplified for permutation-invariant systems, i.e., systems composed of $N$ interacting units whose states are invariant under unit permutations. Such systems can be characterized by a density matrix that grows only polynomially with system size, enabling the determination of the system's von Neumann entropy using a recently developed group theory framework~\cite{cavina2024symmetry}. However, this approach is still limited to finite system sizes. In our article, we present a method enabling one to characterize the asymptotic scaling of multipartite mutual information in the stationary state of permutation-invariant open quantum systems that is applicable in the thermodynamic limit of infinite system size. The proposed approach employs the phase-space description of open system dynamics~\cite{merkel2021phase} together with the group theory framework mentioned above. Our final result is a simple formula that enables one to calculate the multipartite correlations using the standard mean-field description of open quantum system dynamics~\cite{alicki1983nonlinear,lee2013unconventional,lee2014dissipative,benatti2016non,benatti2018quantum,
fiorelli2023mean}. The applicability of our theory is demonstrated on a driven-dissipative version of the Lipkin-Meshkov-Glick (LMG) model~\cite{zhu2015synchronization,tucker2018shattered,zhang2024emergent}. Apart from offering methodological tools, our theory allows us to generalize the conclusion previously drawn for classical permutation-invariant systems~\cite{ptaszynski2024dissipation}: The extensive scaling of multipartite mutual information with system size is not robust to arbitrarily small system perturbations in thermodynamic equilibrium (e.g., in ground states) or in nonequilibrium systems relaxing to fixed points of the mean-field dynamics. However, it can be robust in nonequilibrium systems with time-dependent attractors, such as limit cycles.

The paper is organized as follows: In Sec.~\ref{sec:framework} we present our theoretical framework, including the main result, Eq.~\eqref{eq:asympqinfdet}. In Sec.~\ref{sec:lmg} we illustrate our theory with the driven-dissipative LMG model. Section~\ref{sec:robust} discusses the robustness of extensive scaling of mutual information. Finally, in Sec.~\ref{sec:concl}, we present our concluding remarks and outline potential directions for future research. The appendices provide detailed descriptions of the methods used and include some additional discussions.

\section{Theoretical framework} \label{sec:framework}

\subsection{Permutation-invariant dynamics}

Our work focuses on permutation-invariant networks composed of $N$ quantum $d$-level units (qudits) labeled $i$, that is, networks whose dynamics does not change upon permutation of two units. To realize such a scenario, the units need to be identical, and each pair of units needs to interact in the same way. The state of the total system and of the $i$th unit are denoted as $\rho_T$ and $\rho_i$, respectively. The states and operators in the Hilbert space of a single unit are spanned by the basis $\boldsymbol{v} \equiv \{ v_\alpha \}_{\alpha=1}^{d^2-1}$ of traceless operators $v_\alpha$ that form the Lie algebra of SU(d), plus the identity operator $\mathds{1}_d$. We further denote the basis elements of the $i$th unit as $v_\alpha^{(i)}$ and define the global operators $V_\alpha \equiv \sum_{i=1}^N v_\alpha^{(i)}$. The dynamics of permutation-invariant networks is most commonly described using Lindblad equation of the form (we take $\hbar=1$)~\cite{shammah2018open}
\begin{align} \label{lindblad} \nonumber
	d_t \rho_T=&-i[H,\rho_T]+\sum_{m} \frac{\Gamma_{m}}{N} \mathcal{D}[L_{m}](\rho_T) \\ &+\sum_{n} \gamma_{n} \sum_{i=1}^N \mathcal{D}[L_{n}^{(i)}](\rho_T) \,,
\end{align}	
where $\mathcal{D}[A](\rho) \equiv A \rho A^\dagger-\frac{1}{2} \left\{A^\dagger A,\rho \right\}$. Here, the first term on the right-hand side describes the unitary dynamics, with the effective Hamiltonian $H$ expressed solely in terms of global operators $V_\alpha$. The second term describes the global dissipative dynamics, where each jump of type $m$ is associated with the global dissipation rate $\Gamma_m$ and the Lindblad jump operator $L_m$, which is also expressed solely in terms of global operators $V_\alpha$. Finally, the third term describes local dissipation that is identical for every unit, i.e., every jump of type $n$ in the $i$th unit is associated with the same local dissipation rate $\gamma_n$ and identical local jump operator $L_{n}^{(i)}=\sum_\alpha c_{\alpha n} v_\alpha^{(i)}$. We note that more complex forms of the permutation-invariant Lindblad dynamics, for example, involving collective operator-valued transition rates (i.e., the rates depending on the state of the system), have also been considered in the literature~\cite{rotondo2018open,fiorelli2022phase,fiorelli2023mean,fiorelli2024quantum}. The latter kind of description is necessary to describe interaction with finite-temperature baths.

\subsection{Phase-space description}

We further employ the phase-space description of the system dynamics proposed in Ref.~\cite{merkel2021phase}. Within this framework, permutation-invariant states are represented as
\begin{align} \label{eq:prepres}
\rho_T=\int P(\boldsymbol{\xi}) \rho_{\boldsymbol{\xi}}^{\otimes N} d \boldsymbol{\xi} \,,
\end{align}
where $P(\boldsymbol{\xi})$ is the generalized P-representation of the density matrix and $\rho_{\boldsymbol{\xi}}^{\otimes N}$ is a product state of $N$ identical states of individual subsystems. The latter are parameterized by the generalized Bloch vector $\boldsymbol{\xi} \equiv \{ \xi_l\}_{l=1}^{d^2-1}$ as~\cite{bertlmann2008bloch}
\begin{align}
\rho_{{\boldsymbol{\xi}}} =\frac{1}{d} \left(\mathds{1}_d+\boldsymbol{\xi} \cdot \boldsymbol{v} \right) \,.
\end{align}
The evolution of the distribution $P(\boldsymbol{\xi})$, corresponding to Lindblad equation~\eqref{lindblad}, is then described by the partial differential equation~\cite{merkel2021phase}
\begin{align} \label{eq:phasespacedyn}
d_t P(\boldsymbol{\xi})=-\sum_{l=1}^{d^2-1} \frac{\partial}{\partial \xi_l} \left[g_l(\boldsymbol{\xi} ) P(\boldsymbol{\xi}) \right]+O(1/N) \,,
\end{align}
where $g_l(\boldsymbol{\xi} )$ is the $l$th element of the drift vector $\boldsymbol{g}(\boldsymbol{\xi})$ that corresponds to the mean-field evolution of the generalized Bloch vector,
\begin{align} \label{eq:drift}
d_t \boldsymbol{\xi}_t=\boldsymbol{g}(\boldsymbol{\xi}_t) \,.
\end{align}
This drift vector can be obtained using the commonly employed mean-field approach, which has been thoroughly described in the literature~\cite{alicki1983nonlinear,lee2013unconventional,lee2014dissipative,benatti2016non,benatti2018quantum,
fiorelli2023mean}. The terms of order $O(1/N)$ describe fluctuations that vanish in the thermodynamic limit. Though not demonstrated explicitly, we expect that this kind of description can be generalized beyond dynamics given by Eq.~\eqref{lindblad} to the aforementioned systems with collective operator-valued transition rates, as they also admit an exact mean-field description in the thermodynamic limit~\cite{fiorelli2023mean}.

We now aim to determine the stationary state of the system, described by the stationary P-representation $P_\text{ss}(\boldsymbol{\xi} )$. That state corresponds to the long-time asymptotic solution of Eq.~\eqref{eq:phasespacedyn} when the long-time limit $t \rightarrow \infty$ is taken before the thermodynamic limit $N \rightarrow \infty$, so that the system reaches a unique stationary state\footnote{In the opposite regime, when the thermodynamic limit is taken first, the system initialized in the product state $\rho_{\boldsymbol{\xi}_0}^ {\otimes N}$ stays in the time-evolved product state $\rho_{\boldsymbol{\xi}_t}^ {\otimes N}$ at all times~\cite{merkli2012mean}. In particular, the system may never reach a stationary state, but rather exhibit persistent periodic, quasiperiodic, or chaotic dynamics. This non-commutativity of the thermodynamic and long-time limits is known as Keizer's paradox~\cite{keizer1978thermodynamics}.}. We also focus on the situation where the drift dynamics~\eqref{eq:drift} has a unique ergodic attractor. Based on previous results on classical stochastic systems~\cite{dykman1993stationary,vance1996fluctuations,ge2012landscapes, nicolis1992comments, xu1993internal,geysermans1993thermodynamic,geysermans1996particle,gaspard2020stochastic} [described by equations of motion for classical probability density that are analogous in structure to Eq.~\eqref{eq:phasespacedyn}], we conclude that the stationary P-representation asymptotically converges with $N$ to the invariant probability density of the drift dynamics,
\begin{align} \label{eq:invprobden}
\lim_{N \rightarrow \infty} P_\text{ss}(\boldsymbol{\xi} ) = \lim_{\tau \rightarrow \infty} \frac{1}{\tau} \int_0^{\tau} dt \delta(\boldsymbol{\xi}-\boldsymbol{\xi}_t) \,,
\end{align}
where $\delta$ is the Dirac delta and $\boldsymbol{\xi}_t$ is the solution of Eq.~\eqref{eq:drift} for an arbitrary initial state. In fact, such a distribution remains constant when the system evolves according to Eq.~\eqref{eq:phasespacedyn} in the limit $N \rightarrow \infty$. 

When Eq.~\eqref{eq:drift} admits multiple attractors, the stationary distribution $P_\text{ss}(\boldsymbol{\xi} )$ corresponds to a statistical mixture of different attractors. The stationary state is then determined by either the stochastic jump process between attractors that are well separated in the state space~\cite{drummond1989quantum,dykman2007critical,macieszczak2021theory,kewming2022diverging,lee2024real,xiang2025switching} or the diffusion between adjacent attractors~\cite{carmichael1980analytical}. The detailed analysis of such a case goes beyond the scope of this study. However, we will return to this issue when discussing the robustness of extensive scaling of correlations.

\subsection{Multipartite mutual information}

The goal of our theory is to characterize the multipartite correlations in the stationary state of the system. They are quantified by the \textit{multipartite mutual information} defined as~\cite{watanabe1960information,de2018genuine}
\begin{align} \label{eq:mutinfdef}
I_M \equiv \sum_{i=1}^N S(\rho_i)-S(\rho_T) \,,
\end{align}
where $S(\rho) \equiv-\Tr(\rho \ln \rho)$ is the von Neumann entropy. To calculate the system entropy we use the group theory approach from Ref.~\cite{cavina2024symmetry}. This framework employs Schur's lemma to express the density matrix of a permutation-invariant system in the block-diagonal form
\begin{align} \label{eq:schur}
\rho_T= \bigoplus_{\boldsymbol{\lambda}} p_{\boldsymbol{\lambda}} \rho_{\boldsymbol{\lambda}} \,,
\end{align}
where $\bigoplus$ denotes the direct sum of matrices, and the blocks $\rho_{\boldsymbol{\lambda}}$ are defined within different representations $\boldsymbol{\lambda}$, that is, permutation-invariant subspaces of the total Hilbert space. The index $\boldsymbol \lambda = \{\lambda_j \}_{j=1}^d$ is the ordered partition of $N$ into $d$ integers $\lambda_1 \geq \ldots \geq \lambda_d$ that describe the shape of the Young tableau associated with a given representation. For large system sizes and after taking the continuous limit, the total entropy was shown to scale asymptotically as~\cite{cavina2024symmetry}
\begin{align} \label{eq:entrtotclasens}
   \lim_{N \rightarrow \infty} S(\rho_T)/N = \int p(\boldsymbol{x}) s(\boldsymbol{x}) d \boldsymbol{x} \,,
\end{align}
where $p(\boldsymbol{x})=N^d p_{\boldsymbol{\lambda}}$ is the probability density of the normalized partition $\boldsymbol{x} \equiv \boldsymbol{\lambda}/N$ and $s(\boldsymbol{x}) \equiv -\sum_{j=1}^d x_j \ln x_j$ is an intensive entropy function, having the form of the Shannon entropy. Using Eq.~\eqref{eq:prepres}, the former can be expressed as
\begin{align}
p(\boldsymbol{x})= \int P_\text{ss}(\boldsymbol{\xi}) p_{\boldsymbol{\xi}}(\boldsymbol{x}) d\boldsymbol{\xi} \,,
\end{align}
where $p_{\boldsymbol{\xi}}(\boldsymbol{x})$ is the distribution $p(\boldsymbol{x})$ for a given product state $\rho_{\boldsymbol{\xi}}^{\otimes N}$. 
As shown in Ref.~\cite{alicki1988symmetry}, in the large $N$ limit, the distribution $p_{\boldsymbol{\xi}}(\boldsymbol{x})$ converges to the Dirac delta distribution $\delta \left(\boldsymbol{x} - \boldsymbol{e}_{\boldsymbol{\xi}} \right)$, where $\boldsymbol{e}_{\boldsymbol{\xi}}=\{e_{\boldsymbol{\xi},j} \}_{j=1}^d$ is the set of eigenvalues of $\rho_{\boldsymbol{\xi}}$ arranged in decreasing order. From that we get
\begin{align} \label{eq:entrasympprep}
\lim_{N \rightarrow 
\infty} S(\rho_T)/N=\int P_\text{ss}(\boldsymbol{\xi}) s(\boldsymbol{e}_{\boldsymbol{\xi}}) d \boldsymbol{\xi}= \int P_\text{ss}(\boldsymbol{\xi}) S(\rho_{\boldsymbol{\xi}}) d \boldsymbol{\xi} \,,
\end{align}
where we used $s(\boldsymbol{e}_{\boldsymbol{\xi}}) \equiv -\sum_{j=1}^d e_{\boldsymbol{\xi},j} \ln e_{\boldsymbol{\xi},j} =S(\rho_{\boldsymbol{\xi}})$, which follows from the definition of the von Neumann entropy. Using Eq.~\eqref{eq:prepres} we also get
\begin{align} \label{eq:locstateprep}
\rho_\text{i}=\int P_\text{ss}(\boldsymbol{\xi}) \rho_{\boldsymbol{\xi}} d \boldsymbol{\xi} \,.
\end{align}
Given the assumption of the ergodic dynamics, we can insert Eq.~\eqref{eq:invprobden} into Eqs.~\eqref{eq:entrasympprep}--\eqref{eq:locstateprep}. This yields
\begin{align} \label{eq:asympqinfdet}
\lim_{N \rightarrow \infty} \frac{I_M}{N} = S(\overline{\rho_{\boldsymbol{\xi}_t}})-\overline{S(\rho_{\boldsymbol{\xi}_t})} \,,
\end{align}
where $\overline{(\cdot )} \equiv \lim_{\tau \rightarrow \infty} \tau^{-1} \int_0^\tau (\cdot ) dt$ denotes the infinite-time average. This is the main result of our work, which enables one to determine the macroscopic behavior of multipartite mutual information using the deterministic mean-field dynamics~\eqref{eq:drift}. It generalizes a similar result obtained recently in the context of classical stochastic systems~\cite{ptaszynski2024dissipation} to the quantum regime.

Let us now discuss the implications of our result for the scaling of $I_M$ with system size $N$. Due to the concavity of von Neumann entropy $S(\sum_k p_k \rho_k) \geq \sum_k p_k S(\rho_k)$, which is valid for arbitrary density operators $\rho_k$ and $\sum_k p_k=1$, we have $S(\rho_i) \geq S(\rho_T)/N$, with equality holding if and only if $P_\text{ss}(\boldsymbol{\xi})$ is the Dirac delta. Consequently, the multipartite mutual information~\eqref{eq:mutinfdef} scales subextensively with system size (i.e., $\lim_{N \rightarrow \infty} I_M/N = 0$) when $P_\text{ss}(\boldsymbol{\xi}$) is the Dirac delta, which occurs when the drift dynamics relaxes the system to a single fixed point $\boldsymbol{\xi}_t = \text{const.}$ [see Eq.~\eqref{eq:invprobden}]. Instead, it scales extensively (i.e., $\lim_{N \rightarrow \infty} I_M/N > 0$) otherwise, in particular, when the mean-field dynamics relaxes the system to a time-dependent attractor (e.g., limit cycle or chaotic attractor).

\section{Example: LMG model} \label{sec:lmg}

We illustrate our theory on the driven-dissipative version of the LMG model, whose
phase diagram has been recently thoroughly studied in Ref.~\cite{zhang2024emergent}. It consists of $N$ 2-level units coupled by an isotropic $XY$ interaction and placed in the transverse magnetic field. The algebra of a single unit is spanned by the set of Pauli matrices $\boldsymbol{v}=\boldsymbol{\sigma} \equiv \{\sigma_x,\sigma_y,\sigma_z\}$, while the vector $\boldsymbol{\xi}$ corresponds to the standard Bloch vector of magnetization components $\boldsymbol{m} \equiv \{m_x,m_y,m_z\}$. The system dynamics is described by the master equation
\begin{align}
d_t \rho_T=&-i \left[H,\rho_T \right] +\frac{\Gamma}{N} \mathcal{D} \left[V_- \right](\rho_T) +\gamma \sum_{i=1}^N \mathcal{D} \left[\sigma_+^{(i)} \right](\rho_T) \,,
\end{align}
where
\begin{align} \label{eq:hamlmg}
&H \equiv \frac{\mathcal{J}}{4N} \left( V_x V_x+V_y V_y \right)+\frac{h}{2} V_x \,,
\end{align}
$\mathcal{J}$ is the exchange interaction, $h$ is the transverse magnetic field, $\Gamma$ is the global dissipation rate, $\gamma$ is the local pumping rate, $V_\alpha=\sum_{i=1}^N \sigma_\alpha^{(i)}$ are the global Pauli operators, while $V_\pm=(V_x \pm i V_y)/2$ and $\sigma^{(i)}_\pm=(\sigma_x^{(i)} \pm i \sigma_y^{(i)})/2$ are the global and local ladder operators, respectively. We note that experimental realizations of this model based on cold atoms in optical cavities~\cite{zhu2015synchronization} or trapped ions~\cite{bermudez2013dissipation,shankar2017steady} have been proposed. In particular, unitary and global dissipation terms of this model have been realized in state-of-the-art experiments with cold atoms~\cite{norcia2018cavity,muniz2020exploring}. The local pumping can be realized via a combination of coherent optical excitation from the spin-down state to an adiabatically-eliminated excited state and fast spontaneous decay from that state to the spin-up state, as thoroughly described for the well-known superradiant laser model~\cite{meiser2009prospects}.

The mean-field equations for this model read~\cite{zhang2024emergent}
 \begin{align} \nonumber
d_t m_x&=\mathcal{J} m_y m_z+\Gamma m_x m_z/2 -\gamma m_x/2 \,, \\
d_t m_y &=-\mathcal{J} m_x m_z-h m_z+\Gamma m_y m_z/2-\gamma m_y/2 \,, \\ \nonumber
d_t m_z &=h m_y-\Gamma (m_x^2+m_y^2)/2+\gamma(1-m_z) \,.
\end{align}
These equations can be solved analytically for $h=0$. For $\Gamma<\gamma$, the equations have a stable fixed point at $m_z=1$ and $m_x=m_y=0$. As discussed above, this leads to the vanishing of intensive mutual information $I_M/N$. Instead, for $\Gamma>\gamma$, the equations have a limit cycle solution with (up to arbitrary phase)
\begin{align}
m_x^\text{lc} = m_{xy}^\text{lc} \sin(\omega t) \,, \; m_y^\text{lc} = m_{xy}^\text{lc} \cos(\omega t) \,, \; m_z^\text{lc}=\gamma/\Gamma \,,
\end{align}
where $m_{xy}^\text{lc}=\Gamma^{-1} \sqrt{2 \gamma(\Gamma-\gamma)}$ and $\omega=\mathcal{J}\gamma/\Gamma$. The formation of the limit cycle may be regarded as an instance of many-body synchronization, in which spins synchronize their precession around the $z$-axis~\cite{zhu2015synchronization,shankar2017steady,tucker2018shattered}. Using Eq.~\eqref{eq:asympqinfdet}, the intensive mutual information scales then asymptotically as
 \begin{align}
 \lim_{N \rightarrow \infty} \frac{I_M}{N} = b \left(m_z^\text{lc} \right)-b\left (\sqrt{(m_z^\text{lc})^2+(m_{xy}^\text{lc})^2} \right) \,,
 \end{align}
 where $b(x)=-\sum_{\pm} [(1\pm x)/2] \ln [(1\pm x)/2]$.
For $h \neq 0$, the mutual information can be determined by numerical integration of the mean-field equations.

We also compare the predictions of our theory with the master equation results for finite system sizes. The stationary state of the system is determined using the method from Ref.~\cite{shammah2018open}, while information-theoretic quantities are calculated using the formalism from Ref.~\cite{cavina2024symmetry}. The details of the applied methods are presented in the Appendix~\ref{app:master}. Most importantly, for $h=0$, the steady-state density matrix is diagonal in the Dicke basis, and the number of its nonzero elements is of the order $O(N^2)$. This enables us to consider system sizes up to $N=1600$. Instead, for $h \neq 0$, the number of nonzero density matrix elements is of the order $O(N^3)$, which limits our simulations to smaller system sizes $N \leq 320$.

\begin{figure}
    \centering
    \includegraphics[width=0.9\linewidth]{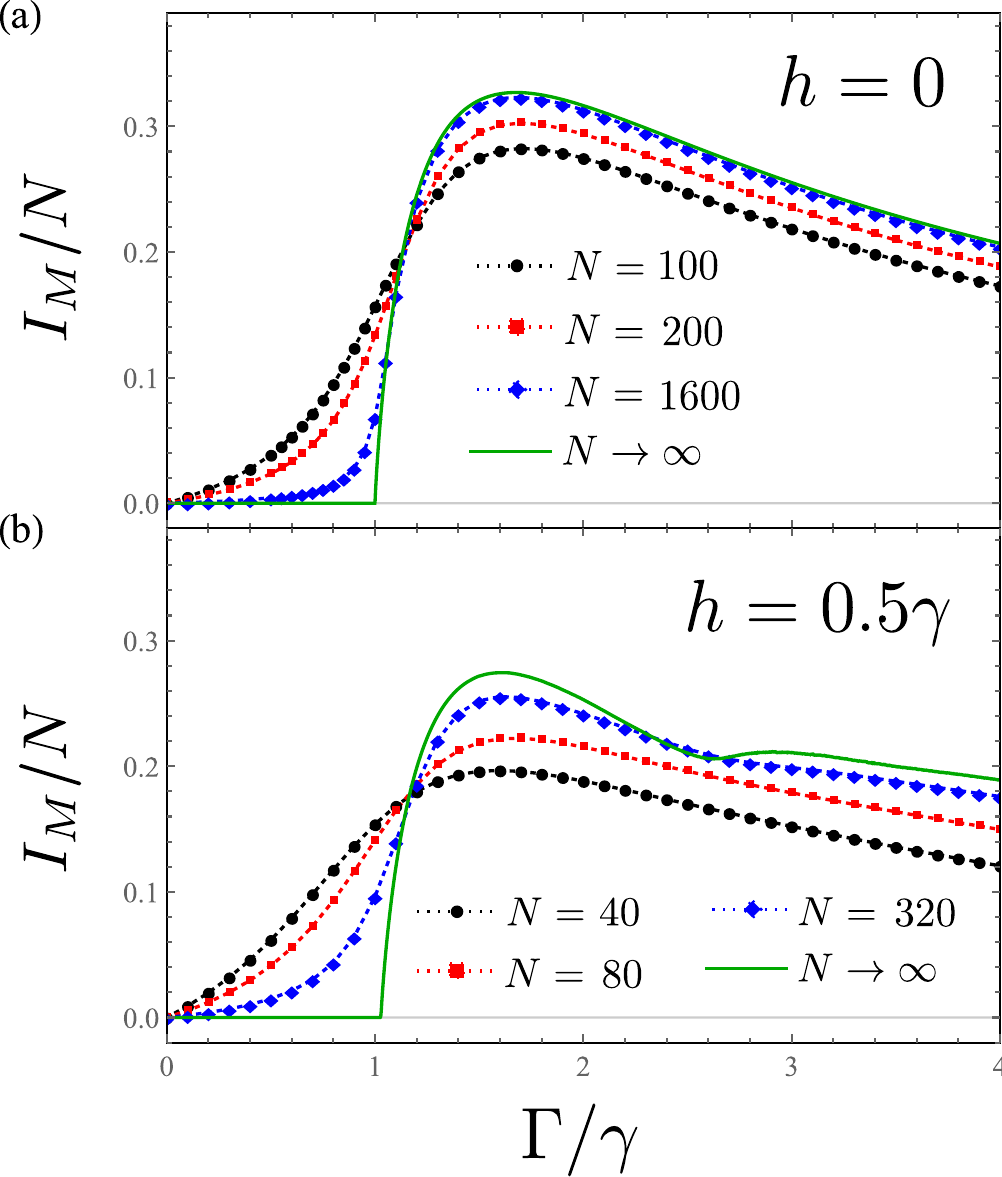}
    \caption{The intensive multipartite mutual information $I_M/N$ in the driven-dissipative LMG model for $\mathcal{J}=3\gamma$ and (a) $h=0$, (b) $h=0.5 \gamma$. The solid green line represents the predictions of our theory and dots represent the master equation results for different system sizes $N$. Dotted lines are added for eye guidance.
    }
    \label{fig:dissipatice}
\end{figure}

In Fig.~\ref{fig:dissipatice}, we present the results for $h=0$ (a) and $h \neq 0$ (b). In the former case, recall, our theory predicts the vanishing of intensive mutual information $I_M/N$ in the fixed-point phase for $\Gamma<\gamma$. In the limit cycle phase for $\Gamma >\gamma$, it instead takes a finite value and exhibits a nonmonotonic behavior, reaching maximum at $\Gamma \approx 1.68 \gamma$. A similar behavior is observed for $h \neq 0$. The master equation results qualitatively reproduce the predictions of our theory: for $\Gamma \lessapprox \gamma$, $I_M/N$ decreases with the system size $N$, gradually approaching 0, while for $\Gamma \gtrapprox \gamma$ it increases with $N$, gradually approaching the theoretical predictions. In particular, for $h=0$ and the large system size $N=1600$, $I_M/N$ is very close to the theoretical predictions in the limit cycle phase. We also notice that for $h \neq 0$, our theory results show a notable dip around $\Gamma \approx 2.5\gamma$, which is not visible in finite-size data. As we discuss in the Appendix~\ref{app:origin}, it appears to result from a strong peak in transverse magnetization $\sqrt{\langle m_{x} \rangle^2+\langle m_{y} \rangle^2}$, which develops very slowly with $N$.

\section{Robustness of extensive scaling} \label{sec:robust}

\begin{figure}
    \centering
    \includegraphics[width=0.9\linewidth]{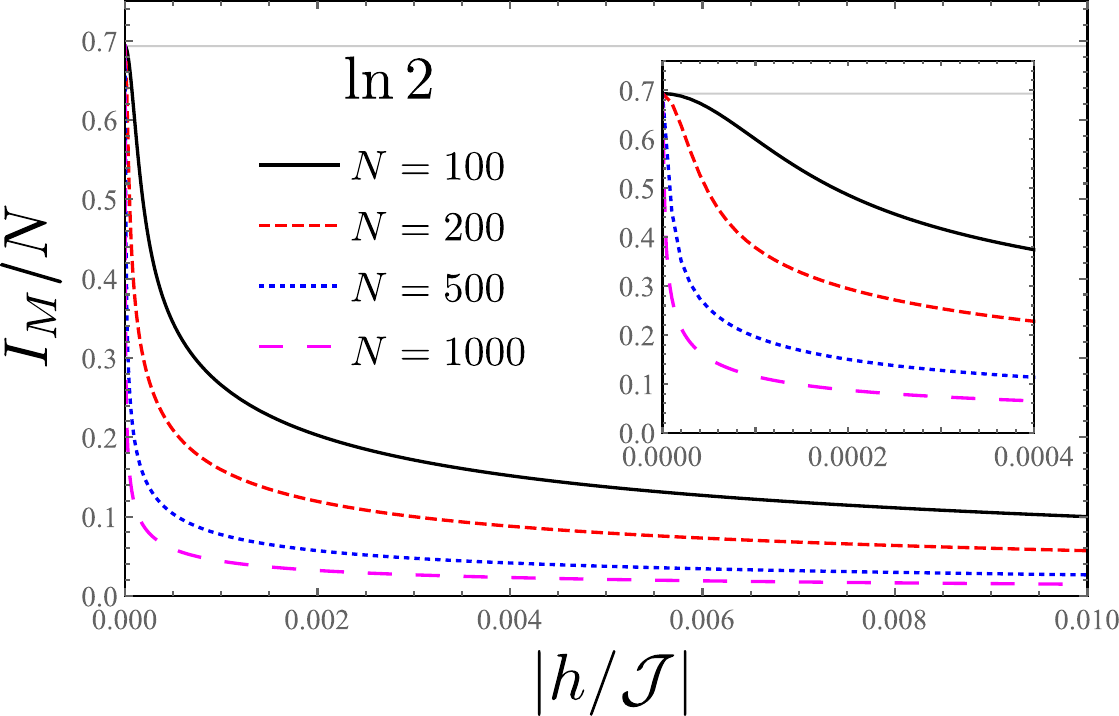}
    \caption{The intensive multipartite mutual information $I_M/N$ in the ground state of the ferromagnetic ($\mathcal{J}<0$) LMG model for different system sizes $N$. The inset presents a smaller range of $h$. Calculation details are presented in Appendix~\ref{app:ground}.
    }
    \label{fig:ground}
\end{figure}

As discussed below Eq.~\eqref{eq:asympqinfdet}, the multipartite mutual information $I_M$ scales extensively if and only if the asymptotic stationary distribution $P_\text{ss}(\boldsymbol{\xi})$ is not the Dirac delta distribution. In the case considered thus far, where the drift dynamics~\eqref{eq:drift} has a unique attractor, this occurs only when this attractor is time-dependent. We now note that when Eq.~\eqref{eq:drift} admits multiple attractors, this is also possible when the stationary state corresponds to a statistical mixture of different fixed points occupied with a finite probability. However, as previously discussed for classical stochastic systems~\cite{ptaszynski2024dissipation}, such situation occurs only when the system dynamics [determined by unitary and dissipative terms of Eq.~\eqref{lindblad}] is perfectly tuned. Otherwise, even when the system admits multiple fixed points, for $N \rightarrow \infty$ it usually tends to relax to a single most-likely fixed point (occupied with probability 1) via stochastic jumps from other fixed points corresponding to metastable states (a phenomenon sometimes called \textit{quantum activation})~\cite{drummond1989quantum,macieszczak2021theory,kewming2022diverging,dykman2007critical,lee2024real,xiang2025switching}\footnote{Recall that in our work we take the long-time limit $t \rightarrow \infty$ before the thermodynamic limit $N \rightarrow \infty$, so that the system relaxes to a unique stationary state. For the opposite order of limits, every fixed point is a stationary state.}. For example, when the system has two fixed points $s \in \{1,2\}$, the switching rate from the point $s$ to $s'$ is proportional to $e^{-N \Phi_s}$, where $\Phi_s$ is the  escape barrier from the state $s$~\cite{xiang2025switching}. Consequently, the steady-state ratio of the occupancies $p_s$ of the fixed points scales as $p_s/p_{s'} \propto e^{N(\Phi_s-\Phi_{s'})}$. As a result, for $N \rightarrow \infty$, the system tends to occupy with probability $1$ the system with the smaller escape rate (i.e., larger $\Phi_s$), unless dynamics of the system is perfectly tuned so that $\Phi_1=\Phi_2$. Apart from such well-tuned case, $I_M/N$ asymptotically vanishes ($\lim_{N \rightarrow \infty} I_M/N=0)$. This means that the extensive scaling of $I_M$ is not robust to small perturbations of unitary or dissipative terms of the system dynamics that break degeneracy of the escape barriers.

This further implies that the extensive scaling of mutual information is not robust in ground or thermal states, which can be regarded as fixed points of thermalizing dynamics (here we assume that time-dependent attractors do not occur in thermal equilibrium, which has been rigorously proven for certain classes of classical~\cite{FalascoReview} and quantum~\cite{watanabe2015absence} systems). We note that ground-state multipartite correlations have attracted some attention~\cite{lucke2014detecting,lourenco2020genuine,lourenço2024genuine,vesperini2024entanglement}. To illustrate that on a particular example, let us consider correlations in the ground state of the Hamiltonian~\eqref{eq:hamlmg}. For ferromagnetic interaction $\mathcal{J}<0$, vanishing transverse field $h=0$, and even $N$, the Hamiltonian has a unique ground state, namely Dicke state~\cite{dicke1954coherence} $V_-^{N/2} (|1 \rangle^{\otimes N})/\sqrt{N!}$, where $|1 \rangle$ is the eigenstate of $\sigma_z$ corresponding to the eigenvalue $1$. In the thermodynamic limit, it can be represented as a uniform mixture of product states $\rho_{\boldsymbol{m}}^{\otimes N}$ with $m_x^2+m_y^2=1$, $m_z=0$. For such a state, $\rho_i=\mathds{1}_2/2$, and thus the intensive mutual information takes a finite value $I_M/N=\ln 2$. However, for any finite transverse field $h$ that breaks the system symmetry, the ground state converges in the thermodynamic limit to a single product state $\rho_{\boldsymbol{m}}^{\otimes N}$ with $\boldsymbol{m}=\{ -\sgn (h),0,0 \}$~\cite{huang2018symmetry}, so that $I_M/N$ vanishes. This behavior is illustrated for finite system sizes in Fig.~\ref{fig:ground}. As shown, $I_M/N$ rapidly decays with $|h|$, the decay being faster for larger $N$, witnessing its vanishing in the thermodynamic limit.

In contrast, in systems that exhibit time-dependent attractors, their existence--and thus the extensive scaling of $I_M$-- can be robust to perturbations. This is illustrated by Fig.~\ref{fig:dissipatice}~(b), where the extensive scaling of $I_M$ is robust to the presence of the transverse field. In fact, the robustness of certain classes of time-dependent attractors (encompassing the system considered) to small perturbations, called \textit{structural stability}, has been rigorously proven within dynamical systems theory~\cite{thompson2002nonlinear,anishchenko2014deterministic, guckenheimer1979structural,tucker1999lorenz,gonchenko2021wild,karateskaia2025robust}; see Appendix~\ref{app:robustness} for details.

\section{Concluding remarks} \label{sec:concl}

Our article demonstrates the power of the phase-space approach~\cite{merkel2021phase} employed together with the group theory framework~\cite{cavina2024symmetry} to describe entropy and information in macroscopic permutation-invariant open quantum systems. While here we focus on steady-state correlations, the proposed method can also be applied in other contexts, including dynamical scenarios. In particular, it can be used to generalize the framework of macroscopic stochastic thermodynamics~\cite{FalascoReview} to quantum systems, e.g., to describe entropy production during quenches between equilibrium states. For the latter goal, the use of thermodynamically consistent master equations~\cite{soret2022thermodynamic}, capable of describing the interaction with finite-temperature reservoirs, would be necessary. We further note that bipartite mutual information has been proposed as a universal order parameter of synchronization between two oscillators, independent of the microscopic details of the considered setup~\cite{ameri2015mutual}. Accordingly, our approach enables one to quantify multipartite synchronization in permutation-invariant networks of oscillators, which have recently received significant attention in both classical~\cite{wood2006critical,wood2006universality,wood2007effects,wood2007continuous,assis2011infinite,assis2012collective,escaff2016synchronization,jorg2017stochastic,zhang2020energy,herpich2018collective,herpich2019universality,meibohm2024minimum,meibohm2024small,guislain2023nonequilibrium,gusilain2024discontinuous} and quantum~\cite{xu2014synchronization, zhu2015synchronization, shankar2017steady, tucker2018shattered, nadolny2024macroscopic} contexts. Finally, we emphasize that our analysis of robustness of extensive scaling is limited to permutation-invariant systems. The open question is whether our conclusions can be generalized to finite-range interacting~\cite{zhang2024emergent,passarelli2022dissipative,mattes2025long,russo2025quantum,wang2025boundary} or disordered~\cite{tucker2018shattered,song2025dissipation} models, where many-body synchronized phases (reminiscent of limit cycles) have also been observed.

\begin{acknowledgments}
K.P.\ and M.C.\ acknowledge the financial support of the National Science Centre, Poland, under the project No.\ 2023/51/D/ST3/01203.
\end{acknowledgments}

\section*{Data availability}
The data and source code that support the findings of this article are openly
available at~\cite{zenodo,github}.

\appendix

\section{Master equation} \label{app:master}
Here, we briefly describe the application of the approach from Ref.~\cite{shammah2018open} to the LMG model considered. This method employs the fact that the permutation invariant state of $N$ spins-1/2 can be expressed as [see Eq.~\eqref{eq:schur}]
\begin{align}
\rho_T=\bigoplus_J \left( \rho_J \otimes \mathds{1}_{\text{dim}_J}/\text{dim}_J \right) \,,
\end{align}
where $J \in \{N/2 \mod 1,\ldots,N/2-1,N/2 \}$ are the eigenvalues of the total angular momentum, and 
\begin{align}
\text{dim}_J=(2J+1)\frac{N!}{\left( \tfrac{N}{2}+J+1 \right)!\left( \tfrac{N}{2}-J\right)!}
\end{align}
is the dimension of the permutation-invariant subspace with given $J$. The elements of $(2J+1) \times (2J+1)$ matrices $\rho_J$ are denoted as $\rho_{J,J_z,J_z'}$ with $J_z,J_z' \in \{ -J,-J+1,\ldots,J \}$. The density matrix can then be effectively represented by the vector $|\rho_T \rrangle$ 
whose $[J(2J+1)(2J-1)/3+(2J+1)(J+J_z)+J+J_z'+1]$th element corresponds to $\rho_{J,J_z,J_z'}$. The steady state is given by the stationary solution of the master equation,
\begin{align} \label{eq:masteqvectorized}
d_t |\rho_T \rrangle = \mathcal{L} |\rho_T \rrangle=0  \,,
\end{align}
where $\mathcal{L} = \bigoplus_J \mathcal{L}_J+\mathcal{L}_\text{loc}$. Here, the matrices $\mathcal{L}_J$ describe the global evolution that preserves the total angular momentum $J$. They can be expressed as~\cite{machnes2014surprising, amshallem2015approaches, uzdin2016speed,minganti2018spectral}
\begin{align} 
	&\mathcal{L}_J =-i \left( \mathds{1}_{2J+1} \otimes \hat{H}_J-\hat{H}_J^T \otimes \mathds{1}_{2J+1} \right) \\ \nonumber &+\frac{\Gamma}{N} \left[\hat{J}_-^* \otimes \hat{J}_- - \frac{1}{2}\mathds{1}_{2J+1} \otimes \hat{J}_-^\dagger \hat{J}_- - \frac{1}{2} \left(\hat{J}_-^\dagger \hat{J}_- \right)^T \otimes \mathds{1}_{2J+1} \right] \,,
\end{align}
where $^*$ denotes the complex conjugate, $^\dagger$ denotes the Hermitian conjugate, $\hat{H}_J=\mathcal{J} (\hat{J}_x^2+\hat{J}_y^2)/N+h \hat{J}_x$, $\hat{J}_\pm = \hat{J}_x \pm i \hat{J}_y$, and $\hat{J}_{x,y,z}$ are spin-$J$ operators. The matrix $\mathcal{L}_\text{loc}$ describes the local dissipation. It is defined such that the generated evolution $d_t |\rho_T \rrangle=\mathcal{L}_\text{loc} |\rho_T \rrangle$ corresponds to the equations of motion
\begin{align} \nonumber
d_t \rho_{J,J_z,J_z'} =& \sum_{k=-1}^1 R^{(k)}_{J+k,J_z-1,J_z'-1} \rho_{J+k,J_z-1,J_z'-1}  \\
&-\frac{\gamma}{2} (N-J_z-J_z') \rho_{J,J_z,J_z'} \,,
\end{align}
where
\begin{subequations}
\begin{align} 
R^{(1)}_{J,J_z,J_z'} &= \frac{\gamma}{2} B_+^{J,J_z} B_+^{J,J_z'} \frac{\tfrac{N}{2}+J+1}{J(2J+1)} \,, \\
R^{(0)}_{J,J_z,J_z'} &= \frac{\gamma}{2} A_+^{J,J_z} A_+^{J,J_z'} \frac{\tfrac{N}{2}+1}{J(J+1)} \,, \\ 
R^{(-1)}_{J,J_z,J_z'} &= \frac{\gamma}{2} D_+^{J,J_z} D_+^{J,J_z'} \frac{\tfrac{N}{2}-J}{(J+1)(2J+1)} \,,
\end{align}
\end{subequations}
with 
\begin{subequations}
\begin{align}
&A_+^{J,J_z}=\sqrt{(J - J_z)(J + J_z+1)} \,, \\
&B_+^{J,J_z}=\sqrt{(J-J_z)(J-J_z-1)} \,, \\ &D_+^{J,J_z}=-\sqrt{(J+J_z+1)(J+J_z+2)} \,.
\end{align}
\end{subequations}
Numerically, the steady state is determined by solving the normal equation $\mathcal{L}^\dagger \mathcal{L} |\rho_T \rrangle =0$, which is obtained by multiplying both sides of Eq.~\eqref{eq:masteqvectorized} by $\mathcal{L}^\dagger$, using the conjugate gradient method described in Refs.~\cite{hestenes1952methods,shewchuk1994introduction}. The entropy of the total system is calculated as~\cite{cavina2024symmetry}
 \begin{align}
 S(\rho_T) = \sum_{J} p_{J} \left[-\ln p_J+S(\bar{\rho}_J)+ \ln \text{dim}_J \right] \,,
 \end{align}
where $p_J=\Tr(\rho_J)$ and $\bar{\rho}_J=\rho_J/p_J$. The local entropies $S(\rho_i)$ are calculated using
 \begin{align} \label{eq:locstate}
 &\rho_i =\tfrac{1}{2} \left(\mathds{1}_2+\langle m_x \rangle \sigma_x+\langle m_y \rangle \sigma_y+\langle m_z \rangle \sigma_z \right) 
 \end{align}
with
\begin{align}
  &\langle m_{x,y,z} \rangle =\frac{2}{N} \sum_{J} \Tr \left (\rho_J \hat{J}_{x,y,z} \right) \,.
  \end{align}

For $h=0$, the stationary matrices $\rho_J$ are diagonal. The diagonal elements $p_{J,J_z}=\rho_{J,J_z,J_z}$ are given by the stationary solution of the rate equations
\begin{align} \nonumber
d_t p_{J,J_z}=&-C_1 p_{J,J_z} + C_2 p_{J,J_z+1}+C_3 p_{J+1,J_z-1} \\
&+C_4 p_{J,J_z-1}+C_5 p_{J-1,J_z-1} \,,
\end{align}
where
\begin{subequations}
\begin{align}
C_1 &=\tfrac{\Gamma}{N} (1+J-J_z)(J+J_z)+ \gamma \left(\tfrac{N}{2} -J_z\right) \,, \\
C_2 &=\tfrac{\Gamma}{N} (J-J_z)(J+J_z+1) \,, \\
C_3 &= \gamma \frac{(J-J_z+1)(J-J_z+2)(\tfrac{N}{2}+J+2)}{2(J+1)(2J+3)} \,, \\
C_4 &= \gamma \frac{(J-J_z+1)(J+J_z)(\tfrac{N}{2}+1)}{2J(J+1)} \,, \\
C_5 &= \gamma \frac{(J+J_z-1)(J+J_z)(\tfrac{N}{2}-J+1)}{2J(2J-1)} \,.
\end{align}
\end{subequations}
The total entropy can then be calculated as
\begin{align}
S(\rho_T)=\sum_{J,J_z} p_{J,J_z} \left(-\ln p_{J,J_z}+\ln \text{dim}_J \right) \,,
\end{align}
while the local states are given by Eq.~\eqref{eq:locstate} with $\langle m_x \rangle=\langle m_y \rangle=0$ and $\langle m_z\rangle=(2/N) \sum_{J,J_z} J_z p_{J,J_z}$.

 \section{Origin of dip in $I_M/N$ for $h \neq 0$} \label{app:origin}

%
\begin{figure}
	\centering
	\includegraphics[width=0.95\linewidth]{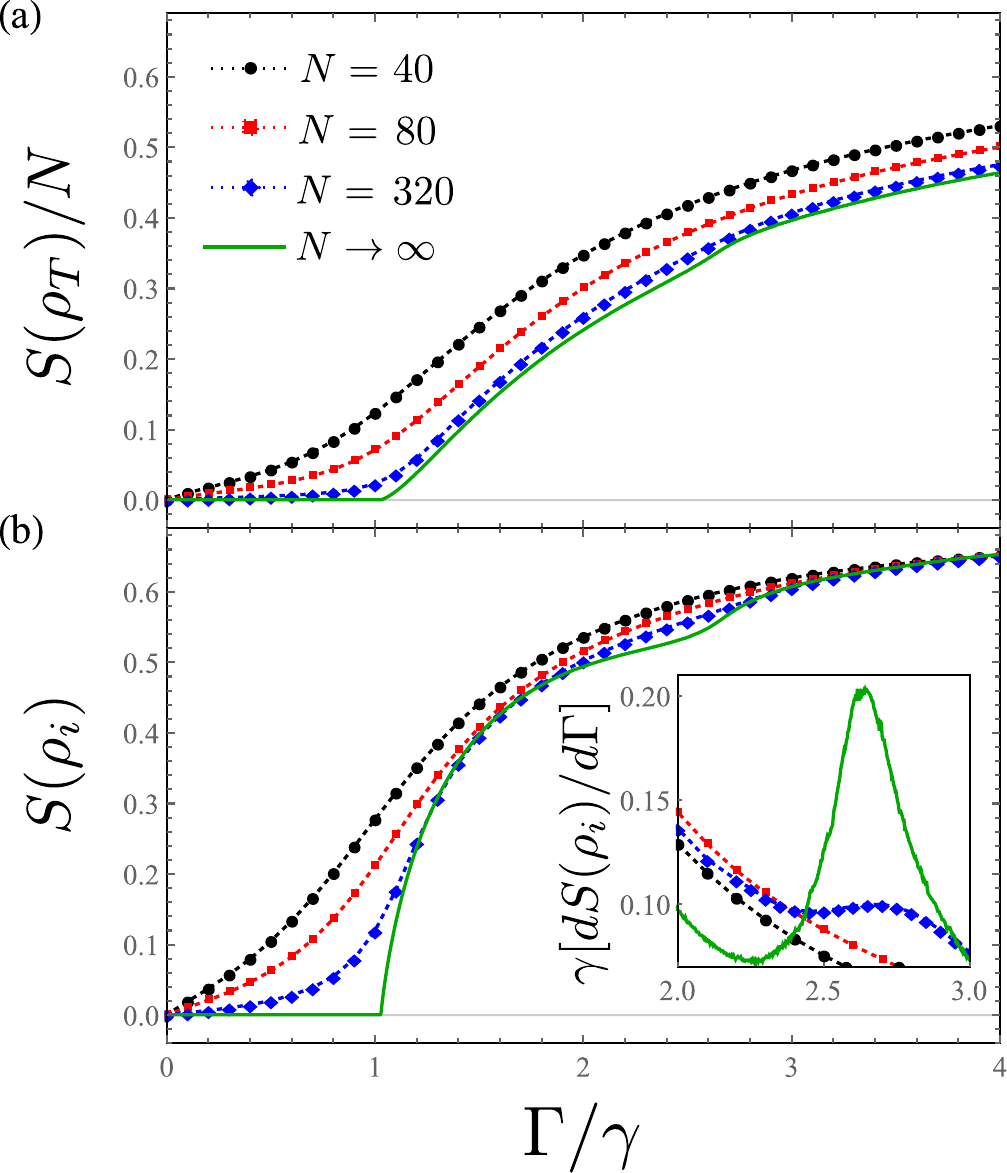}		
	\caption{(a) Scaled total entropy $S(\rho_T)/N$ and (b) local entropy of a single unit $S(\rho_i)$ as a function of $\Gamma/\gamma$. The inset in (b) shows the derivative $dS(\rho_i)/d\Gamma$ for a smaller range of $\Gamma/\gamma$. Parameters and notations as in Fig.~1(b) in the main text. For $N \rightarrow \infty$, we apply our theory as $S(\rho_T)/N = \overline{S(\rho_{\boldsymbol{m}_t})}$, $S(\rho_i)=S(\overline{\rho_{\boldsymbol{m}_t}})$.}
	\label{fig:stot-sloc}
\end{figure}
%
%

In Fig.~1(b) in the main text, presenting the behavior of multipartite mutual information in the driven-dissipative LMG model for $h =0.5\gamma$, we observed the presence of a notable dip in our theoretical predictions for $I_M/N$ that occurs around $\Gamma/\gamma \approx 2.5$, which is not visible in the finite-size results. Here we elaborate on the origin of this dip and its apparent absence for finite system sizes $N$. To do this, we first consider the behavior of two individual components of $I_M/N$, that is, $S(\rho_T)/N$ and $S(\rho_i)$ [see Eq.~(7) in the main text]. It is presented in Fig.~\ref{fig:stot-sloc}. As shown, the scaled total entropy $S(\rho_T)/N$ gradually converges from above to the predictions of our theory, which is related to the subextensive scaling of the first two terms in Eq.~(A8) in the Appendix A; see Ref.~\cite{cavina2024symmetry} for details. For large $N=320$, we observe a very good quantitative agreement between our theory and finite-size results. For local entropy $S(\rho_i)$, in the limit cycle phase ($\Gamma/\gamma \gtrapprox 1.02$), the predictions of our theory are also close to the finite-size results for large $N=320$. However, they show a minor hump that occurs in the same region as the dip in $I_M/N$ (around $\Gamma/\gamma \approx 2.5$), and--like this dip--is also not clearly visible for finite system sizes in the plot of $S(\rho_i)$. However, traces of its presence become visible in the derivative $dS(\rho_i)/d\Gamma$ for large $N=320$; see the inset of Fig.~\ref{fig:stot-sloc}(b).

%
\begin{figure}
	\centering
	\includegraphics[width=0.95\linewidth]{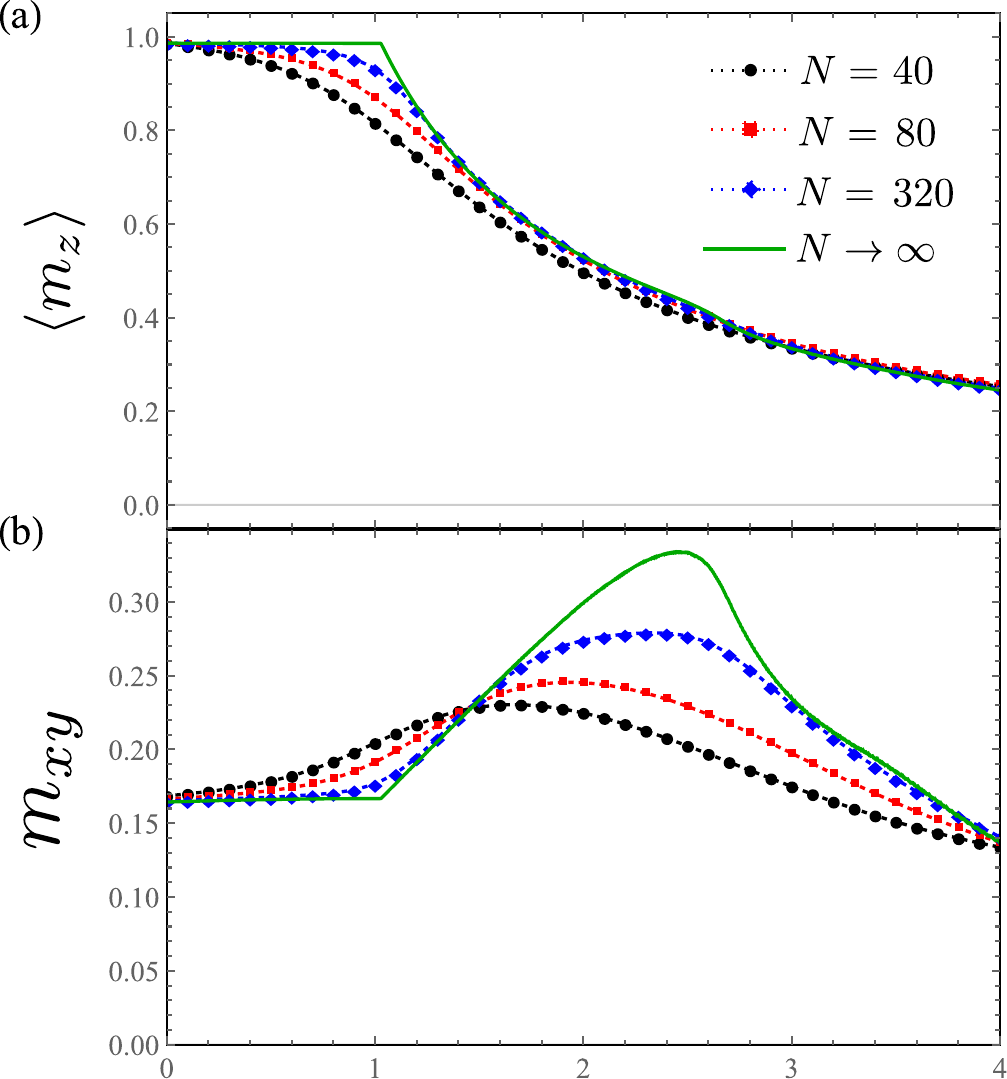}		
	\caption{(a) Longitudinal magnetization $\langle m_z \rangle$ and (b) transverse magnetization $m_{xy}=\sqrt{\langle m_{x} \rangle^2+\langle m_{y} \rangle^2}$ as a function of $\Gamma/\gamma$ (note different scales on $y$ axis). Parameters and notations as in Fig.~1(b) in the main text. For $N \rightarrow \infty$, we apply the mean-field theory as $\langle m_\alpha \rangle=\overline{(\boldsymbol{m}_{t})_\alpha}$.}
	\label{fig:magn}
\end{figure}
%
%

To gain insight into the origin of this hump in $S(\rho_i)$, and the resulting dip in $I_M/N$, we now consider the behavior of the magnetization components $\langle m_{x,y,z} \rangle$, which determine the local entropy through Eq.~(A9) in the Appendix A. It is presented in Fig.~\ref{fig:magn}. First, in mean-field results ($N \rightarrow \infty$) for the limit cycle phase, longitudinal magnetization $\langle m_z \rangle$ exhibits monotonic decay, with only a minor hump observed around $\Gamma/\gamma \approx 2.5$. The finite-size results for large $N$ are mostly very close to the mean-field behavior. 

Much more notable is the behavior of transverse magnetization $m_{xy}=\sqrt{\langle m_{x} \rangle^2+\langle m_{y} \rangle^2}$. In the mean-field results, it exhibits a very pronounced peak around $\Gamma/\gamma \approx 2.5$, where the dip in $I_M/N$ is observed. In fact, the enhancement of $m_{xy}$ makes the local states $\rho_i$ more pure, reducing the local entropy $S(\rho_i)$. This explains the origin of the hump in $S(\rho_i)$, and thus in $I_M/N$. At the same time, the finite-size results exhibit a gradual, but very slow, emergence of this peak with increasing $N$. This slow development of the peak in transverse magnetization rationalizes why we are unable to observe the resulting dip in $I_M/N$ in finite-size results. 

Finally, we notice that the slow development of the peak in $m_{xy}$ could possibly be related to the proximity of the considered region of the phase diagram ($\Gamma/\gamma \approx 2.5$, $h=0.5\gamma$) to the region, where the dynamics becomes chaotic via a period-doubling bifurcation (for higher values of $h$); see Fig.~2 in Ref.~\cite{zhang2024emergent}. As previously reported, in the chaotic regime, the strong enhancement of fluctuations by chaos may cause a strong deviation of the stationary state properties of the system from mean-field predictions even for relatively large system sizes~\cite{fox1990effect, fox1991amplification, keizer1992reply, wang1997intrinsic, wang1998master}. A detailed exploration of the system behavior in the chaotic regime is beyond the scope of this study.

 \section{Ground-state mutual information} \label{app:ground}
To calculate the mutual information in the ground state of the ferromagnetic LMG model, we first note that the ground state is pure, so $S(\rho_T)=0$. To determine the local entropies $S(\rho_i)$, we use the fact that the ground state is characterized by the maximum value of the angular momentum $J=N/2$~\cite{huang2018symmetry}. Consequently, the local states $\rho_i$ are given by Eq.~\eqref{eq:locstate} with
 \begin{align}
 \langle m_{x,y,z} \rangle=2\langle g| \hat{J}_{x,y,z}| g \rangle/N \,,
 \end{align}
 where $|g \rangle$ is the ground state of the Hamiltonian $\hat{H}_J=\mathcal{J} (\hat{J}_x^2+\hat{J}_y^2)/N+h \hat{J}_x$ with $J=N/2$.

 \section{Robustness of time-dependent attractors} \label{app:robustness}
As discussed in the main text, the existence of time-dependent attractors of the mean-field dynamics is a sufficient condition of extensive scaling of the mutual information $I_M$. Consequently, the extensive scaling of $I_M$ is robust to perturbations of the system dynamics when the existence of time-dependent attractors is also robust. The latter property is known as \textit{structural stability} in dynamical systems theory, where it is defined as robustness of the attractor existence to small perturbations of the drift vector $\boldsymbol{g}(\boldsymbol{\xi})$ [that can result from perturbation of the unitary or dissipative part of Eq.~\eqref{lindblad}]. This property has been proven for several classes of chaotic attractors~\cite{guckenheimer1979structural,tucker1999lorenz,gonchenko2021wild,karateskaia2025robust}, as well as for a certain class of periodic orbits, called hyperbolic limit cycles~\cite{thompson2002nonlinear,anishchenko2014deterministic}. To verify whether the limit cycle is hyperbolic, one needs to consider the Jacobian of the mean-field dynamics evaluated around the time-evolved state $\boldsymbol{\xi}_t$,
\begin{align}
\mathbb{J}(\boldsymbol{\xi}_t)=\left[ \partial_{\xi_l} g_k (\boldsymbol{\xi}_t) \right ]_{1 \leq k,l \leq d^2-1} \,.
\end{align}
We focus on a situation where the system exhibits a time-periodic limit cycle attractor with $\boldsymbol{\xi}_{t}=\boldsymbol{\xi}_{t+T}$, where $T$ is the oscillation period, and consider the time evolution of the fundamental matrix $\mathbb{M}(t)$,
\begin{align}
d_t \mathbb{M}(t)=\mathbb{J}(\boldsymbol{\xi}_t) \mathbb{M}(t) \,,
\end{align}
with $\mathbb{M}(0)=\mathds{1}$. The hyperbolicity of the limit cycle is then determined by the eigenvalues of the monodromy matrix $\mathbb{M}(T)$, called the \textit{Floquet multipliers}: the cycle is hyperbolic if a single Floquet multiplier equals $1$, while all others have moduli different from 1. In particular, for attractive limit cycles, the latter multipliers have moduli smaller than 1. 

We can now verify that this property is satisfied for the limit cycle of the LMG model considered in the main text, which guarantees the robustness of the extensive scaling of $I_M$. For comparison, we note that periodic orbits in the cooperative resonance fluorescence model~\cite{carmichael1980analytical}, which recently gained significant attention as a paradigmatic example of a driven-dissipative time crystal~\cite{iemini2018boundary}, are not hyperbolic. Consequently, the periodic oscillations in the latter setup are not robust, but rather are suppressed by arbitrarily small local dephasing~\cite{shammah2018open} or local dissipation~\cite{song2025dissipation}.

\bibliography{bibliography}	
	
\end{document}